\documentstyle[preprint,aps]{revtex}
\newcommand{\be}{\begin{equation}}
\newcommand{\ee}{\end{equation}}
\newcommand{\la}{\langle}
\newcommand{\ra}{\rangle}
\newcommand{\bea}{\begin{eqnarray}}
\newcommand{\eea}{\end{eqnarray}}

\newcommand{\pa}{\partial}

\newcommand{\br}{{\bf r}}

\begin{document}
\baselineskip 4.5ex
\draft
\title{\bf Prospect of creating a composite 
fermi/bose superfluid}
\author{ 
Eddy Timmermans,$^{1}$ Kyoko Furuya,$^{2}$ Peter W. Milonni,$^{1}$ \\
and Arthur K. Kerman$^{3}$\\
 {\small $^{1}$ Theoretical Division (T-4), Los Alamos National Laboratory,} 
{\small Los Alamos, NM 87545} \\
{\small $^{2}$ Universidade Estadual de Campinas, Instituto de
Fisica `Gleb Wataghin'} \\
{\small c.p. 6165, 13083-970, Campinas, SP, Brazil} \\
{\small $^{3}$ Center for Theoretical Physics, Laboratory for Nuclear Science
 and} \\
{\small Department of Physics, Massachusetts Institute of Technology,}\\
{\small Cambridge, MA 02139}}
\maketitle
\begin{abstract}
We show that composite fermi/bose superfluids can be created 
in cold-atom traps by employing a Feshbach
resonance or coherent photoassociation. The bosonic molecular condensate 
created in this way implies a new fermion pairing mechanism
associated with the exchange of fermion pairs between
the molecular condensate and an atomic fermion
superfluid.  We predict macroscopically coherent, Josephson-like
oscillations of the atomic and molecular populations in response
to a sudden change of the molecular energy, and suggest that these 
oscillations will provide an experimental signature of the pairing.

\end{abstract}

\pacs{PACS numbers(s):03.75.Fi, 05.30.Jp, 32.80Pj, 67.90.+z}


	The cold-atom technology that has led to the observation
of dilute-gas Bose-Einstein condensates (BEC's) \cite{becs} enables
the creation and study of novel superfluids having a degree
of flexibility that is highly unusual in traditional low
temperature physics.  In addition to the observed 
boson gas superfluids, the prospect of studying fermion superfluidity
in the gas phase provides a powerful motivation \cite{leggett} for the
atom cooling efforts with fermion isotopes \cite{Jin}.  
S-wave Cooper pairing -- the
most likely mechanism for achieving fermion superfluidity in
the low-density gases -- relies on the mutual attraction of
two distinguishable types of fermions \cite{stoof}.  
In this Letter, we show that some of the techniques that are
being developed \cite{exp} to increase the mutual attraction and, hence, the
critical temperature for Cooper-pairing, create a fundamentally
novel superfluid state.  Specifically, the interaction-altering
schemes that involve a coherent atom-molecule coupling such 
as the Feshbach resonance \cite{feshbach1} and coherent 
photoassociation \cite{photo} create
a condensate of bosonic molecules mutually coherent with the
atomic Cooper-paired gas.  In treating the dynamics of this
system we note an interesting experimental signature: the system
responds to a sudden change of the single-molecule energy by means 
of macroscopically coherent, Josephson-like oscillations of the
atomic and molecular populations.

	Here we discuss the effects on a mixture of $\pm 1$ fermion atoms
of a Feshbach resonance that coherently converts $\pm$ atom pairs
to molecules $m$.  Restricting ourselves to the relevant s-wave processes,
the coupling is described by the term
\begin{equation}
H_{a-m}=\int d^3\br\ \; \alpha\left[\hat{\psi}_m^{\dag}\hat{\psi}_{+1}\hat{\psi}_{-1}
 \ + {\rm h.c.}\right] 
\label{e3}
\end{equation}
in the Hamiltonian.  Earlier \cite{feshbach1}, 
we had pointed out that such
interaction can produce mutually coherent atom-molecule boson
superfluids.  In the fermion gas mixture, we now show that
this interaction coherently couples 
an atomic paired fermion superfluid to a BEC of $m$-molecules.
We characterize boson superfluidity 
in the broken symmetry description by the appearance of a complex-valued
order parameter, the condensate field $\phi$: $\phi({\bf r}) = \langle
\hat{\psi} ({\bf r}) \rangle$, where $\hat{\psi}$ is the bosonic annihilation
field operator and ${\bf r}$ the position.  
Similarly, we characterize s-wave pairing of fermionic atoms in
internal states $\pm 1$ by a pairing-field order
parameter $\langle \hat{\psi}_{\sigma} ({\bf r}) \hat{\psi}_{-\sigma} 
({\bf r}) \rangle
= F_{\sigma} ({\bf r}) \neq 0$, where $\sigma = \pm 1$. 
It is not difficult to see that the interaction 
(\ref{e3}) implies the presence of
a bose condensate of molecules ($\la\hat{\psi}_m\ra\neq 0$) if the atoms are
fermi-paired  ($\la\hat{\psi}_{+1}\hat{\psi}_{-1}\ra\neq 0$). The equation
of motion for $\la\hat{\psi}_m\ra=\phi_m$, obtained by taking the expectation
value of the Heisenberg equation $i\hbar\pa\hat{\psi}_m/\pa
t=[\hat{\psi}_m,\hat{H}]$, contains a source term
$\alpha\la\hat{\psi}_{+1}\hat{\psi}_{-1}\ra\neq 0$, implying that $\phi_m$ cannot remain zero.
Moreover, as we will show, the interaction (1) plays a more fundamental
role than in the BEC analogue: the atom-molecule interaction
{\it causes} the superfluidity and provides a new pairing mechanism.

	From a theoretical point of view, however, this mechanism
is not entirely new as the Hamiltonian (1) has been proposed as
a phenomenological model for high $T_{c}$ superconductivity \cite{fbmodel}.
However, experimental investigations have cast serious
doubt on the validity of the so-called fermion-boson model to
explain high $T_{c}$ superconductivity.  It is then important 
to realize that, in contrast to the phenomenological description of 
high $T_{c}$ superconductors,
the atomic gases are described by a first principle's theory with
parameters that are calculated and measured unambiguously.  We also
note that the trapped atom systems have
a different and much wider range of experimentally controllable
variables: the external potentials, inter-particle scattering lengths,
fermion densities, relative numbers of unlike fermions,
the strength ($\alpha$) of the atom-molecule coupling and the
value of the single molecule energy can all be varied.  In particular,
the real time variation of the single molecule energy suggests
testing the mutual long-range coherence by observing Josephson-like 
oscillations in the atomic and molecular populations.  Prompted by this 
exciting possibility
we develop a time-dependent mean-field description of the dynamics, 
different from the usual statistical treatments of the fermion-boson model.

	We start with the 
Heisenberg equations for the field operators, $\hat{\psi}_m$ and
$\hat{\psi}_{\pm 1}$. As we will justify a posteriori,
the scattering length interactions can be neglected and we obtain
\begin{eqnarray}
i\hbar\dot{\hat{\psi}}_{+1}=\left[-{\hbar^2\nabla^2\over
2m}+V_{+1}+\epsilon_{+1}\right]\hat{\psi}_{+1}-
\alpha\hat{\psi}_m\hat{\psi}_{-1}^{\dag}
\ , 
\nonumber \\
i\hbar\dot{\hat{\psi}}_{-1}=\left[-{\hbar^2\nabla^2\over
2m}+V_{-1}+\epsilon_{-1}\right]\hat{\psi}_{-1}+
\alpha\hat{\psi}_m\hat{\psi}_{+1}^{\dag}
\ ,
\nonumber \\
i\hbar\dot{\hat{\psi}}_{m}=\left[-{\hbar^2\nabla^2\over
4m}+V_{m}+\epsilon_{m}\right]\hat{\psi}_{m}+
\alpha\hat{\psi}_{+1}
\hat{\psi}_{-1}
\ , 
\label{e:4}
\end{eqnarray}
where $V_{\pm 1}$ and $V_m$ denote the external potentials
and $\epsilon_m$ and $\epsilon_{\pm 1}$ the internal energies.
The fact that fermion atoms of different spin projection can have
significantly different energies in the external field of
magnetic traps (e.g., $\epsilon_{-1} << \epsilon_{+1}$) does
not necessarily prevent their pairing. Though the conversion of
atoms from the $+1$ to the $-1$ state would lower the
energy, the necessary spin-flip interaction may be negligible,
as demonstrated by the simultaneous trapping of essentially stable
populations of atoms in non-degenerate internal
states \cite{Ref1}.  Such systems reach
a quasi-equilibrium, characterized, in this case,
by fixed fermion numbers: a total of $N_{+}$ ($+1$)--atoms
and $N_{-}$ ($-1$)--atoms, with respective chemical potentials
$\mu_{+}$ and $\mu_{-}$.  Chemical equilibrium further
dictates that $\mu_{m} = \mu_{+}
+ \mu_{-}$, where $\mu_{m}$ is the molecular chemical potential.  
A simple transformation of the field operators,
$\hat{\psi}_{\sigma}' = \hat{\psi}_{\sigma} 
\exp\left[ (i/\hbar) \int^{t} \{\epsilon_{\sigma}(\tau)+\mu_{\sigma}\} d\tau
\right]$ and $\hat{\psi}_{m}'=\hat{\psi}_{m} \exp\left[ (i/\hbar)
\int^{t} \{\epsilon_{+}(\tau)+\epsilon_{-}(\tau)+\mu_{m}\} d\tau\right]$,
illustrates that the dynamics  does not depend
on the fermion energy difference, $\epsilon_{+} - \epsilon_{-}$,
but only on the molecular excitation energy $\epsilon = \epsilon_{m} -
[\epsilon_{+}+\epsilon_{-}]$.
Indeed, the transformed operators satisfy
\begin{eqnarray}
i \hbar \dot{\hat{\psi}}_{\sigma} &=&
\hat{h}_{\sigma}
\hat{\psi}_{\sigma} - \sigma \alpha \hat{\psi}_{m}
\hat{\psi}_{-\sigma}^{\dag}  ,
\nonumber \\
i \hbar \dot{\hat{\psi}}_{m} &=&
 \hat{h}_{m}
\hat{\psi}_{m} + \; \alpha \hat{\psi}_{+} \hat{\psi}_{-} \;  ,
\label{e:5}
\end{eqnarray}
where we drop the primes and define $\hat{h}_{\sigma} =
\left[ - \frac{\hbar^{2} \nabla^{2}}{2m} + V_{\sigma} - \mu_{\sigma} 
\right]$,
$ \hat{h}_{m} = \left[ - \frac{\hbar^{2} \nabla^{2}}{4m} + V_{m} + 
\epsilon -\mu_{m} \right]$. 
As in Cooper-pairing \cite{pairingDG}, the annihilation of a fermion
is accompanied by the creation of a fermion of opposite
``spin" \cite{spins}.
Accordingly, the
quasiparticle state of quantum number $j$ and spin
$\sigma$, $|j,\sigma\rangle$, is a mixture of a $\sigma$ particle and
a $-\sigma$ hole.  We characterize the fermion pairing field by the
particle and hole amplitudes for bringing the many-body state
$|0\rangle$ to the excited $|j,\sigma \rangle$-- state:
 $u_{j,\sigma} ({\bf r},t) = 
\langle 0 | \hat{\psi}_{\sigma}({\bf r},t) | j,\sigma \rangle$,
$ v_{j,\sigma} ({\bf r},t) = - \sigma
\langle 0 | \hat{\psi}^{\dagger}_{\sigma} ({\bf r},t) | j , -\sigma 
\rangle$.  We obtain the zero temperature equations of motion by taking 
the excitation amplitudes of the fermion Heisenberg
equations and the expectation value of the $\dot{\hat{\psi}}_{m}$
equation.  In the assumption of molecular condensate coherence, 
$\langle \hat{\psi}_{m} \rangle \approx 
\phi_{m}$, and $\langle 0| \hat{\psi}_{m}^{\dagger} \hat{\psi}_{\sigma}
| j, \sigma \rangle \approx \phi_{m}^{\ast} u_{j,\sigma}$
\cite{Gaussian}, this mean-field procedure gives 
Bogoliubov de-Gennes (BDG)-like equations for the $u/v$-parameters
coupled to a Gross-Pitaevskii (GP)-like
equation for $\phi_{m}$:
\begin{eqnarray}
i \hbar \frac{d}{dt} \left( \begin{array}{c}
u_{j,\sigma} \\ v_{j,-\sigma}
\end{array}  \right) 
\; \; &=& 
\left( \begin{array}{cc}
\; \hat{h}_{\sigma} \; \; \; & -\alpha \phi_{m} \\
\; -\alpha \phi_{m}^{\ast} \; \; \; & - \hat{h}_{-\sigma}
\end{array} \right) \; 
\left( \begin{array}{c}
u_{j,\sigma} \\ v_{j,-\sigma}
\end{array} \; ,
\right) \;, \; \sigma = \pm 1 \;  ,
\nonumber \\
i \hbar \frac{\partial \phi_{m} }{\partial t} \; \; \; &=& \; \;
\hat{h}_{m} \phi_{m} + \alpha F_{+1}
\; , 
\label{e:6}
\end{eqnarray}
where 
$F_{+1} = \langle \hat{\psi}_{+1} \hat{\psi}_{-1} \rangle \;$
$= \sum_{j}
\langle 0| \hat{\psi}_{+1} |j,+1\rangle
\langle j,+1|\hat{\psi}_{-1}|0\rangle =
\sum_{j} u_{j,+} v_{j,-}^{\ast}$
represents the pairing field.
The anticommutator
relations satisfied by the $\hat{\psi}_{\sigma}$-fields imply that
$F_{+1} = - F_{-1}$, and that $\sum_{j}[u_{j,\sigma}({\bf r})
u_{j,\sigma}^{\ast}({\bf r}') + v_{j,\sigma} ({\bf r})
v_{j,\sigma}^{\ast} ({\bf r}')] = \delta ({\bf r} - {\bf r}')$.

	In a homogeneous system, $V_{+} = V_{-} = 0$, 
contained in a macroscopic
volume $\Omega$, the excited many-body states 
have definite momentum, $|j,\sigma\rangle \rightarrow 
|{\bf k},\sigma\rangle$
and the $u_{j,\sigma}, v_{j,\sigma}$ mode
functions are plane waves, $u_{j,\sigma}({\bf r},t)
= u_{k,\sigma}(t) \exp(i {\bf k} \cdot {\bf r})/\sqrt{\Omega}$. 
Fermion anticommutation implies $|u_{k,\sigma}|^{2} + |v_{k,\sigma}|^{2} = 1$,
and the $\hat{h}_{\sigma}$-operators of (\ref{e:6}) give real numbers,
$\hat{h}_{\sigma} \rightarrow h_{k,\sigma} = [\hbar^{2} k^{2}/2m
- \mu_{\sigma}]$.
For $N_{+} = N_{-} = N$, the ideal situation
to achieve superfluidity, $\mu_{+} = \mu_{-} = \mu$, 
the $\sigma =  \pm 1$ equations are identical, 
\begin{eqnarray}
i \hbar \dot{u}_{k} &=& h_{k} u_{k} - \alpha \phi_{m} v_{k}
\nonumber \\
i \hbar \dot{v}_{k} &=& - \alpha \phi_{m}^{\ast} u_{k} - h_{k} v_{k}
\nonumber \\
i \hbar \dot{\phi}_{m} &=& (\epsilon - 2 \mu) \phi_{m} + 
\alpha F_{+} \; ,
\label{e:7}
\end{eqnarray}
where $F_{+} = \sum_{\bf k} u_{k} v_{k}^{\ast}/\Omega$
and we have dropped the spin subscripts.  The single
molecule energy, $\epsilon$, can be varied experimentally
by altering the
magnetic field strength in the Feshbach resonance, or the
laser detuning in the case of coherent photoassociation.

\underline{Statics}
In the static ($\dot{\phi}_{m} = 0$) off-resonant limit
where $\epsilon$ greatly exceeds all other single particle
energies including the fermi-energy $\epsilon_{F}
= \hbar^{2} k_{F}^{2}/2m$, where $k_{F}$ denotes the
Fermi-momentum, $k_{F} = (6 \pi^{2} n)^{1/3}$ and $n = N/\Omega$,
few molecules are formed and $\mu \approx \epsilon_{F}$.
We recover the BDG equation
from (\ref{e:6}) by substituting $\phi_{m}$ from
the GP-equation, $\phi_{m} 
= - (\alpha F_{+1}) / (\epsilon - 2 \mu)$, in 
the $u/v$-equations which gives
gives the usual Fermi-pairing equations \cite{pairingDG} with an effective
scattering length 
\begin{equation}
a_{\rm eff} = - \left(\frac{m}{4\pi\hbar^{2}} \right) \; \frac{\alpha^{2}}{\epsilon-2 \mu} \; ,
\label{e:n}
\end{equation}
similar to the effective scattering length of
of the binary atom Feshbach resonance \cite{feshbach1}.
Not surprisingly, the off-resonant critical temperature $T_{c}$ obtained
in the finite-temperature mean-field treatment takes on the
form of a negative scattering length $T_{c}$:
$T_{c} \approx T_{c,eff} = 0.6 \exp(-\pi/2 k_{F} |a_{eff}|)$.
Near-resonance, $T_{c}$ follows from coupled integral equations 
\cite{fbmodel}, although our numerical results
for a realistic system (a $Rb_{86}$--mixture at
$n \approx 2. \times 10^{13} cm^{-3}$, $\alpha \sqrt{n} = \epsilon_{F}$)
differed from $T_{c,eff}$ by less than a factor of 2 (giving
near-resonant $T_{c}$'s $\sim 10^{-7}$ K) \cite{press}.

	   The static $u$ and $v$--amplitudes in (\ref{e:7}) evolve
according to the excitation energies $E(k)$ of the
$|{\bf k},\sigma\rangle$--state, $u_{k}(t)
= u_{k} \exp[-iE(k)t/\hbar]$, $v_{k}(t) = v_{k}
\exp[-iE(k)t/\hbar]$.  The resulting $u/v$--equations have a non-trivial 
solution if
\begin{equation}
E(k) =
\sqrt{ \left( \hbar^{2} k^{2} /2m - \mu \right)^{2}
+ |\alpha \phi_{m}|^{2} } \; \; ,
\label{e:8}
\end{equation}
giving a superconductor-like dispersion 
with gap $|\alpha \phi_{m}|$.
Since the gap measures the success of the fermion
pairing in lowering the overall energy, 
Eq. (\ref{e:8}) illustrates that it is the 
mutual coherence of the Bose
and Fermi fields that pairs the fermions,
i.e. the boson and fermion fields maintain macroscopic
coherence by `feeding off' each other.

	The static $u$ and $v$ `coherence' factors
that follow from (\ref{e:7}), $u_{k}^{2} = [1 + h_{k}/E(k)]/2$
and $v_{k}^{2} = [1 - h_{k}/E(k)]/2$, characterize the 
fermion pairing physics.  The occupation number of the 
single fermion ${\bf k}$-state is given by $v_{k}^{2}$
and the Fourier-transform of 
$g(k) = v_{k}/u_{k}$ gives the wavefunction 
$\varphi({\bf r}={\bf r}_{i}-{\bf r}_{j})$ of
the Cooper pairs.  The range of the pair wavefunction,
$\varphi$,
$L_{c}^{2} = \int d^{3} r \; r^{2} \varphi({\bf r})
/ \int d^{3} r \; \varphi({\bf r})$, follows from
the long wavelength behavior of $g(k) \approx g(0) 
[ 1 - (k L_{c})^{2}/6]$, $k \rightarrow 0$.  We find that
$\hbar^{2}/(2mL_{c}^{2}) = E(k=0)/6 = \sqrt{4 \mu^{2} + 
(\alpha \phi_{m})^{2}}/6$ and $L_{c}$  
is largest in the off-resonant limit
where $\mu \approx \epsilon_{F}$ and $|\alpha \phi_{m}|
<< \epsilon_{F}$, so that $L_{c} \approx \sqrt{3} k_{F}^{-1}$.
Typically \cite{trap}, the local value of $k_{F}^{-1}$
exceeds the length scale on which the density 
of a trapped degenerate fermion system varies so
that the homogeneous results also apply to the 
trap system in the sense of a Thomas-Fermi description.

	Provided $|\alpha \phi_{m}| << \mu$, we can
approximate the $F_{+}$--integral analytically and the
properly renormalized \cite{renorm} GP-like equation
yields $\phi_{m} \approx 8 e^{-2} (\mu/\alpha)
\exp\left(-\frac{\pi}{2 k_{F} |a_{eff}|} \right)$,
where $\mu \approx \epsilon_{F}$.  From the
exponential dependence on $(k_{F} a_{eff})^{-1}$ (as
compared to $\phi_{m} \propto \epsilon^{-1}$ in the 
resonant BEC \cite{feshbach1}) it follows that
only close to resonance does a measurable molecular
field appear, $|a_{eff}| \geq 0.4 k_{F}^{-1}$.
This regime is accessible to Feshbach
resonance experiments.  In that case, the limitation
in $\epsilon$ ($a_{eff}$) corresponds to a magnetic
field interval around the resonant value $B_{0}$, $|\delta
B| < \delta {\overline B}$, where $\delta B = B - B_{0}$.
We determine the interval size $\delta {\overline B}$
from $|a_{eff}| \geq 0.4 \; k_{F}^{-1}$ with
\cite{stenger} $a_{eff} = - a_{s} (\Delta/\delta B)$,
where $a_{s}$ denotes the ordinary scattering length
of the unlike fermion interaction
and $\Delta$ represents the measured magnetic field 
`width': $\delta {\overline B} = 2.5 \times (|a_{0}| k_{F})
|\Delta|$.  To obtain representative numbers, we consider
the $Na_{23}$ resonances observed by Stenger et al. \cite{stenger}
for which $|\Delta| = 0.1 G, 1. G$ and $4 G$.  Taking
$n = 10^{12} cm^{-3}$, \cite{stoof} and $a_{s} = 3.3 \; nm$ (the $Na$
triplet scattering length), we find
$\delta {\overline B} = 0.03 \; |\Delta| = 3 mG, 30 mG, 0.12 G$
exceeding current experimental magnetic field resolution (typically
better than $1 mG$).  Also, within $|\delta B| \leq \delta
{\overline B}$,
$|a_{eff}| = |\Delta/\delta B| a_{s} > 30. \; a_{s}$,
which justifies our approximation of
neglecting the $a_{s}$--interactions in (\ref{e:4}).  
As the system is tuned closer to resonance,
$\phi_{m}$ and $\mu$ need to be solved self-consistently
from (\ref{e:7}) and the condition
of particle conservation: $ N = \Omega |\phi_{m}|^{2}
+ \sum_{k} v_{k}^{2}$.

\underline{Dynamics}  As mutually coherent superfluids, 
the atom-molecule system 
responds to a sudden change of $\alpha$ or $\epsilon$ 
by means of macroscopically coherent
Josephson-like atomic and molecular population oscillations. 
In the figure, we show the molecular condensate population
as $\epsilon$ varies in time \cite{renorm2}: at t=$\hbar/\epsilon_{F}$,
$\epsilon$ was suddenly raised from 
$8 \epsilon_{F}$ to $14 \epsilon_{F}$, and
returned to its initial value at $t=1.2 \hbar/\epsilon_{F}$.
The time-unit $\hbar/\epsilon_{F} = 2.1 \; A (n/n_{0})^{-2/3} \mu {\rm sec}$,
where $A$ is the atomic number and $n_{0}$ denotes a reference
density of realistic value, $n_{0} = 10^{12} cm^{-3}$, indicates
time scales that are readily 
accessible to the available technology. 
Note that a population oscillation in a binary atom 
collision would start when the atoms meet,
which happens `randomly' in a classical description of
the atomic center-of-mass
motion, thereby washing out any such
oscillations in a many-body average. 
Thus, macroscopically coherent
atom-molecule oscillations constitute a genuine manifestation of 
mutually coherent long-range order.

	Given the problems encountered in the evaporative
cooling of neutral fermion atoms \cite{Jin}, it may be of
future interest to use coherent molecule formation
in an alternative cooling scheme, such as the example we
now discuss.  (i) We photoassociate finite temperature atoms
to molecules. (ii) We switch off the lasers and cool the bosonic
molecules evaporatively to BEC. (iii) We turn the lasers on
and sweep the detuning (which plays the role of $\epsilon$)
of the resulting fermi/bose superfluid from negative
to positive values.  If we can neglect spontaneous photon
emission and other heating and loss-processes, this
procedure results in an ultra-cold fermi/bose system
of mostly fermionic atoms. (iv) We switch off the lasers.
	
	In conclusion, we have pointed out that
the coherent atom-molecule interactions available
in present-day atom cooling experiments suggest
the exciting prospect of creating a cold atom superfluid
of novel structure, consisting of mutually
coherent field components of different
spin statistics.  Although this novel type
of fermion pairing mechanism
was previously proposed as a phenomenological
model for high $T_{c}$ superconductivity, the
atom trap systems could provide the first realization.
In addition to the advantage of being described by a well-defined,
first principles Hamiltonian, we show that
the added flexibility of controlling the detuning
can be used to diagnose this new long-range coherence.

\newpage
\centerline{\Large Figure}
\noindent
\underline{Figure 1} Plot showing the fraction of atoms
in the molecular condensate (i.e. $|\phi_{m}|^{2}/n$), as 
the fermi/bose system responds to a sudden increase at $t=1$
of the molecule energy from
its initial value, $\epsilon=8 \epsilon_{F}$
to $\epsilon=14 \epsilon_{F}$.
At $t=1.2$,
$\epsilon$ was returned to its initial value.
The energy unit, $\epsilon_{F}$, is the
fermi energy of the atoms and the unit of time on the horizontal
axis is equal to $\hbar/\epsilon_{F}$.
We choose a realistic atom-molecule coupling constant for which
$\alpha \sqrt{n} = \epsilon_{F}$.


\begin{thebibliography}{99}
\bibitem{spins} For convenience we refer to the $\sigma$-labels as
``spins."
\bibitem{press} Similarly, second-order perturbation around
the off-resonant state gives, to leading order, the expected
pressure, $P = (4/5) n \epsilon_{F} + (4 \pi \hbar^{2}/m) a_{eff} n^{2}
= (4/5) n \epsilon_{F} [ 1 + (5/3\pi) (k_{F} a_{eff})]$.  Although 
$a_{eff} < 0$ in the $\epsilon > 0$ off-resonant regime,
$k_{F} |a_{eff}| << 1$, so that the positive fermi pressure,
$(4n \epsilon_{F}/5)$, dominates.
\bibitem{trap} In the middle of 
a spherically symmetric harmonic oscillator trap of frequency
$\omega_{T}$ (corresponding to a single particle
ground state size $l_{T} = \sqrt{\hbar/m\omega_{T}}$),
that contains a Fermi-degenerate system of $N$ indistinguishable fermions,
the local $k_{F}^{-1}$--value equals $k_{F}^{-1}
= \frac{l_{T}}{2 (6 N)^{1/3}}$.  For typical values,
$N \geq 10^{5}$, this length is much shorter than
the extent of the trapped fermion system, $L_{T}
= l_{T} 2 (6 N)^{1/3}$.
\end{thebibliography}
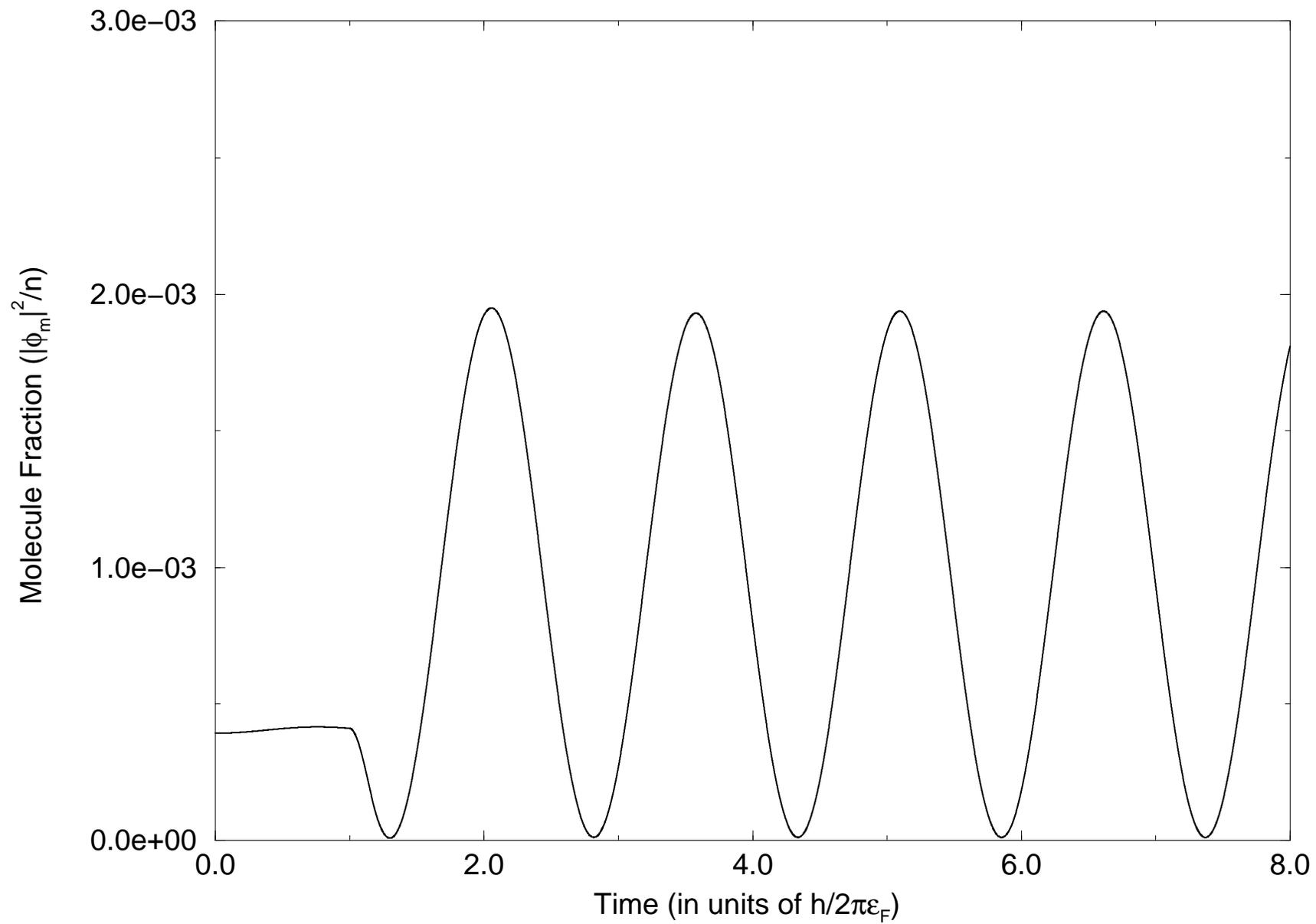
\end{document}